# Technical Note: A Feasibility Study on Deep Learning-Based Radiotherapy Dose Calculation

Yixun Xing, Dan Nguyen, Weiguo Lu, Ming Yang, Steve Jiang[†]
Medical Artificial Intelligence and Automation (MAIA) Laboratory, Department of Radiation Oncology, University of Texas Southwestern Medical Center, Dallas, TX, 75390, USA

[†]Correspondence author. Email: Steve.Jiang@UTSouthwestern.edu

## Abstract

**Purpose:** Various dose calculation algorithms are available for radiation therapy for cancer patients. However, these algorithms are faced with the tradeoff between efficiency and accuracy. The fast algorithms are generally less accurate, while the accurate dose engines are often time consuming. In this work, we try to resolve this dilemma by exploring deep learning (DL) for dose calculation.

**Methods:** We developed a new radiotherapy dose calculation engine based on a modified Hierarchically Densely Connected U-net (HD U-net) model and tested its feasibility with prostate intensity-modulated radiation therapy (IMRT) cases. Mapping from an IMRT fluence map domain to a 3D dose domain requires a deep neural network of complicated architecture and a huge training dataset. To solve this problem, we first project the fluence maps to the dose domain using a modified ray-tracing algorithm, and then we use the HD U-net to map the ray-tracing dose distribution into an accurate dose distribution calculated using a collapsed cone convolution/superposition (CS) algorithm.

**Results:** It takes about one second to compute a 3D dose distribution for a typical 7-field prostate IMRT plan, which can be further reduced to achieve real-time dose calculation by optimizing the network. For all eight testing patients, evaluation with Gamma Index and various clinical goals for IMRT optimization shows that the DL dose distributions are clinically identical to the CS dose distributions.

**Conclusions:** We have shown the feasibility of using DL for calculating radiotherapy dose distribution with high accuracy and efficiency.

## 1. Introduction

Various dose calculation algorithms have been developed for cancer radiotherapy and are available in commercial treatment planning systems (TPSs), ranging from simple pencil beam models to more complicated convolution/superposition algorithms to even more advanced Monte Carlo methods. It is well known that simple dose engines like pencil beam models can be fast but less accurate, while advanced algorithms like Monte Carlo methods can be very accurate but slow. Ahnesjo et al. provided a comprehensive review of various dose calculation algorithms used for external beam photon radiation therapy.[1] Overall, there is a tradeoff between calculation efficiency and accuracy for all dose calculation algorithms. There is a clinical need to develop new dose engines that are both accurate and efficient.

Recently, deep learning (DL) has become a driver of many new real-world applications ranging from language translation[2,3] to computer vision.[4,5] A deep learning architecture, U-net,[6] has been successfully applied to predict dose distributions for prostate cancer radiotherapy.[7] The model was also modified to predict the dose distributions for head and neck cancer patients and lung cancer patients with heterogeneous beam setups.[7-9]

In this work, we explore the feasibility of using DL for accurate and fast radiotherapy dose calculation. Specifically, we test the Hierarchically Densely connected U-net (HD U-net) model for intensity-modulated radiation therapy (IMRT) dose calculation for prostate cancer patients using a low-accuracy, first-order prior as input.

## 2. Methods and Materials

### 2.1. Deep neural network architecture

Because of domain differences, directly mapping from 2D fluence maps to a 3D dose distribution could be challenging for DL, requiring a complicated network structure and a large amount of training data. Projecting the fluence maps first into the dose domain to use as a prior would greatly simplify the model itself and its training. Our idea is to use ray-tracing (RT) dose calculation,[10] which is fast and inexpensive, to obtain a prior from the fluence maps and the patient CT as input for the DL model. In this feasibility study, we use the dose distributions from the collapsed cone convolution/superposition (CS algorithm[11-13] as the desired output dose distribution. The RT algorithm provides inaccurate dose calculation because it does not consider the scatter components from the primary beam. The CS algorithm is considered accurate in this regard and has been widely used in clinical practice.

Thus, the dose calculation problem can be translated into a mapping problem from the RT dose distribution to the CS dose distribution using DL, and we designed a deep neural network to learn the relationship between the RT and the CS dose distributions and to capture the scattering effect. Keep in mind that the DL model can be trained with the output data from any accurate dose calculation algorithm.

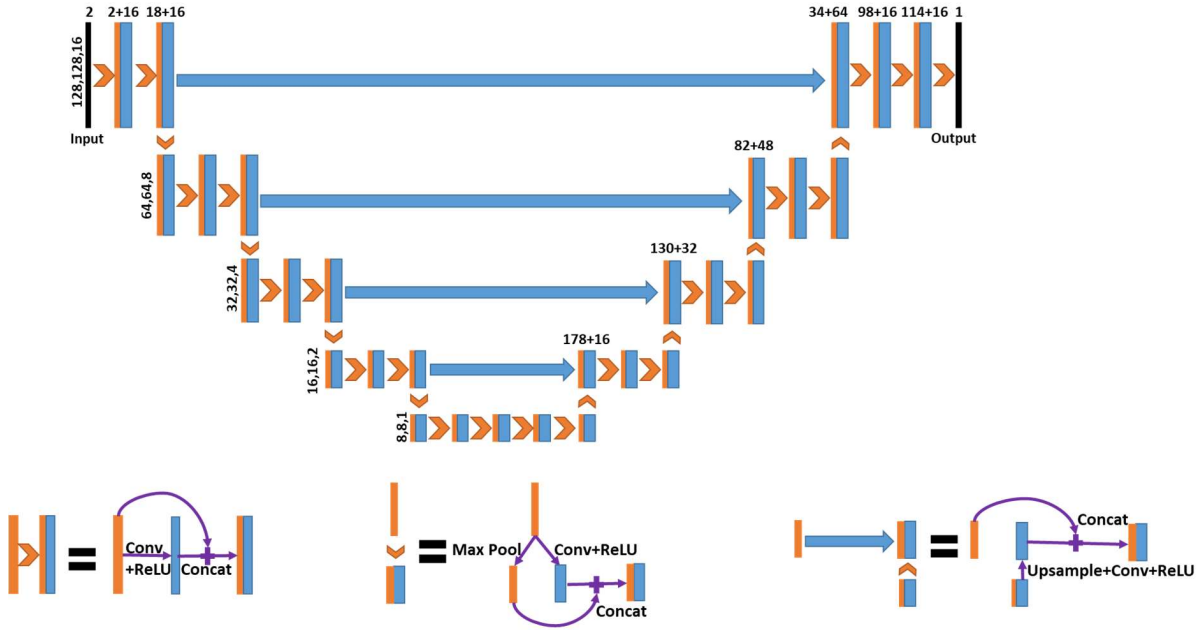

Figure 1. Schematic of the HD U-net architecture used for the dose calculation. The number above the boxes represents the number of concatenated features, and those to the left show the size of the three-dimensional feature maps.[9]

Figure 1 shows the neural network architecture, HD U-net, used in this work. The general HD U-net structure was proposed for three-dimensional dose prediction for head and neck cancer patients.[9] We adopted the main neural network operations of the original HD U-net and made modifications to the architecture. As can be seen in Figure 1, the HD U-net contains five levels to reduce the feature size from 128×128×16 down to 8×8×1, where 2×2×2 max pooling was performed between layers. The convolutional kernel of size 3×3×3 was implemented during convolution with zero padding to maintain the feature size. Sixteen feature maps were used in each convolutional step on the left half of the network. On the right half of the network, the number of filters increased by 16 for each level from the bottom to the top. The convolution and rectified linear unit (ReLU) operations were followed by batch normalization (BN) in the HD U-net, as suggested for dose prediction with U-net.[7] Because we observed no overfitting issue, we kept the dropout rate at 0. The learning rate was carefully adjusted to $10^{-4}$, and the Adam algorithm[14] was selected as the optimizer to minimize the loss function of the mean squared error (MSE). The deep learning model was built and implemented in Keras with Tensorflow[15] as the back end.

## 2.2. Data collection

We used 78 prostate IMRT patients in this feasibility study. For each patient, the treatment plan with seven 6-MV photon beams was re-calculated using both RT and CS algorithms. The input and output volume dimensions were 256×256×64 for 77 patients and 256×256×62 for the remaining one patient.

## 2.3. Training and evaluation

Seventy patients were randomly assigned as the training data, and the remaining eight patients were held aside as separate testing data. A five-fold cross-validation was performed during the training stage to assess the performance stability and variability of the proposed HD U-net model. The 70 patients were divided into 5 folds with 14 patients in each fold. For every training round, four folds (56 patients) were used for training, and the last fold was reserved for validation. The weights were randomly initialized for each round and then updated based on its training dataset. The five-fold training results were collected and assessed to ensure stability before a final model was constructed on the 70 patients combined. The final model was then tested on the other eight patients.

During each training iteration, a patch of size 128×128×16 was randomly selected from the patient CT and dose images. This training-by-patch method, similar to data augmentation, could reduce overfitting. The number of epochs, with around 56 iterations in each epoch, was optimized through fine tuning and determined to be 300.

For the eight testing cases, we computed the gamma passing rate at 1mm/1% and 2mm/2% for the assessment. We also analyzed the statistics, including the mean, standard deviation, minimum, and maximum of the differences between CS and DL dose distributions for various IMRT clinical goals. We also calculated the dose volume histogram (DVH) comparison of CS, RT, and DL dose distributions for testing patients. In addition, we computed the error histograms for critical regions.

The DL model was trained on NVIDIA Tesla K80 dual-GPU cards with 12 GB dedicated RAM. The evaluation on testing patients was performed on one NVIDIA Tesla V100 GPU card with 32 GB dedicated RAM.

## 3. Results

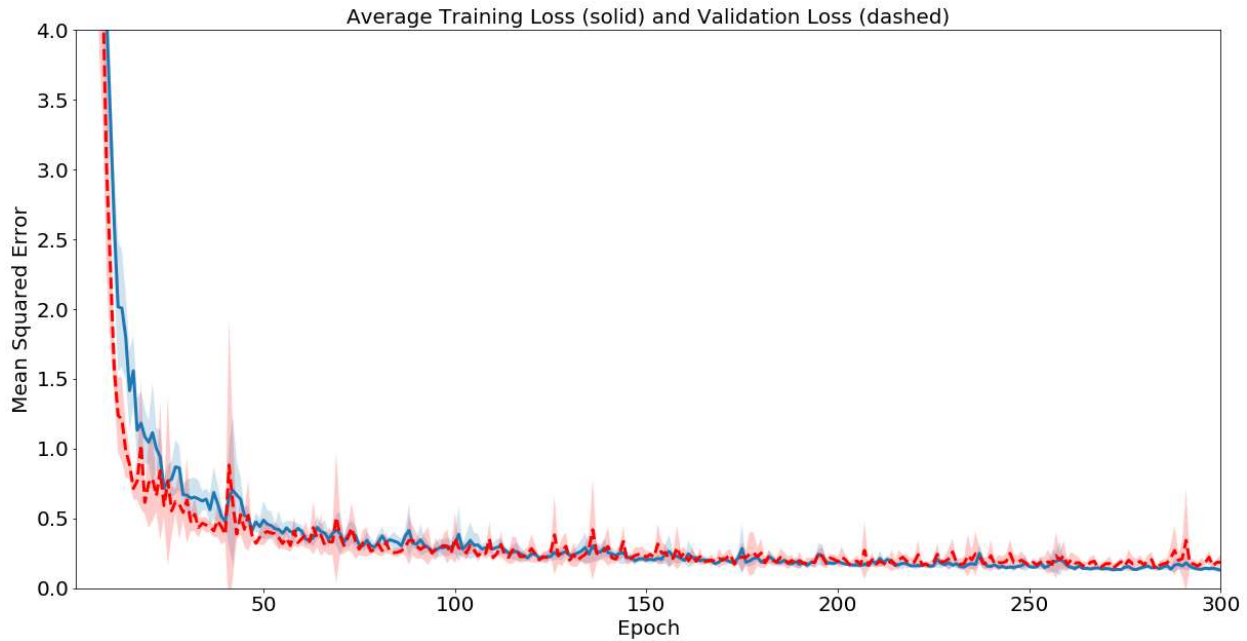

Figure 2. The average loss across the 5 cross-validation folds, where the solid line represents the training loss and the dashed line shows the validation loss.

Figure 2 displays the training and validation losses for the HD U-net. Generally, both the training and validation losses decrease as the epoch increases. Based on all five validation losses, we observed no over-fitting issue.

The trained model was applied to calculate the eight testing plans, and the average calculation time was 1.19 seconds with a standard deviation of 0.01 seconds. Figure 3 illustrates the CS dose (a), the DL dose (b) and their difference (c) distributions loaded on the CT slice for an example patient. As can be seen, the DL dose is very close to the CS dose, with a difference of less than ~2% of the prescription dose. Figure 3 (d) shows the DVH curves for the same patient. The solid lines represent the CS dose, while the dashed lines and the dotted lines represent the DL and RT doses, respectively. One can see that the DL DVH curves are completely covered by the CS DVH curves and can hardly be seen in Figure 3 (d), indicating a clinically acceptable accuracy for the DL dose and a considerable improvement from the RT dose.

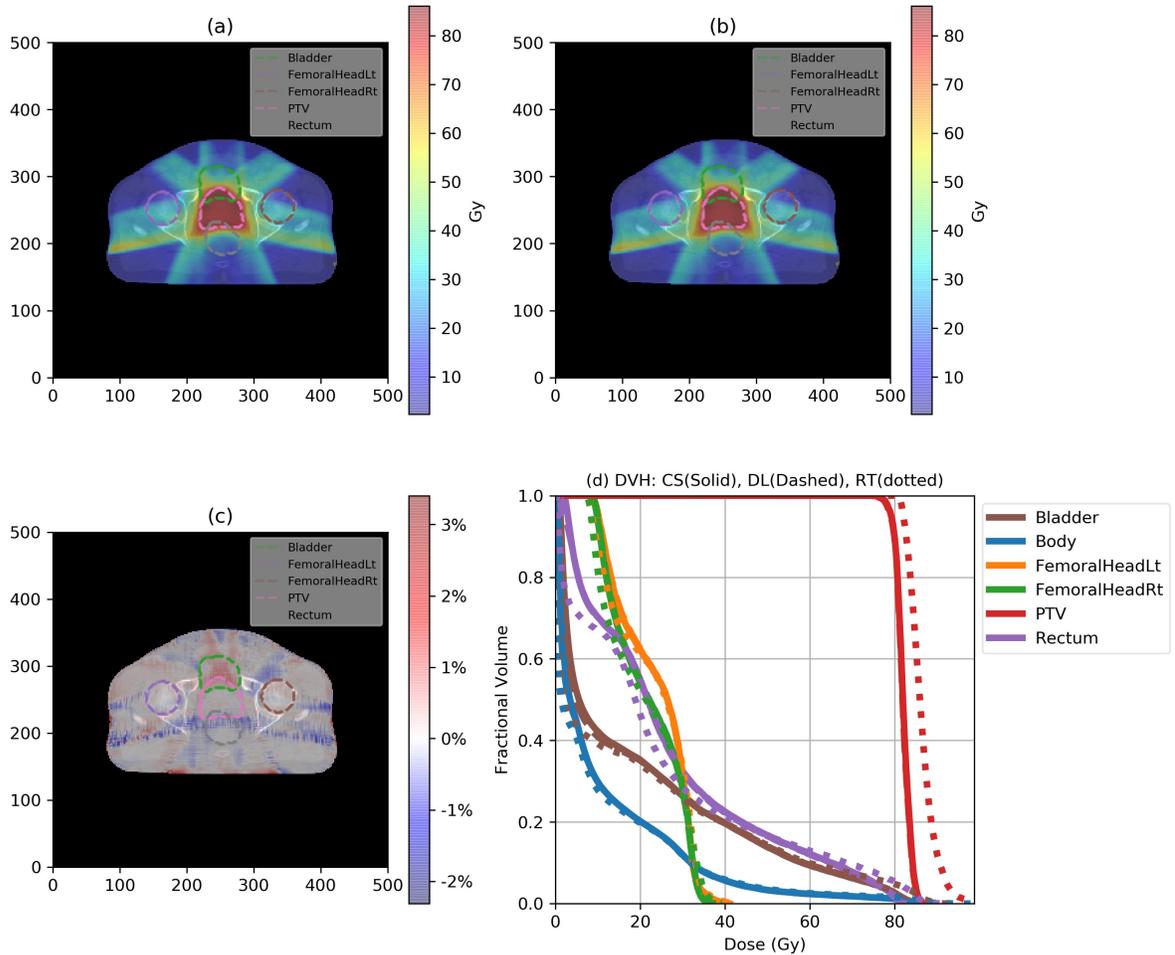

Figure 3. (a) One slice of the CS dose distribution; (b) the corresponding slice of the DL dose distribution; (c) the relative dose difference ($\frac{D_{DL}-D_{CS}}{D_{Rx}}$) distribution; and (d) the DVH plots of the CS (solid), the DL (dashed), and the RT (dotted) doses for one example patient.

Table 1 shows the gamma passing rates at 1mm/1% and 2mm/2% criteria for the DL doses for the eight testing patients, using the CS dose as the reference. As can be seen, the gamma analysis indicates the high accuracy of the DL dose in both settings. The gamma passing rate for the 1mm/1% criterion is as high as 98.50% on average with a standard deviation of 1.6%. For the 2mm/2% criterion, the corresponding numbers become 99.9% and 0.10%. These numbers indicate that there is essentially no clinically meaningful difference between DL and CS dose distributions. Table 2 displays the mean, standard deviation, min, and max of the difference in IMRT optimization objectives between CS and DL dose calculations. The differences of DL and CS algorithms in the $D_{95}$ of PTV for the eight testing patients fall between -1.38 Gy and 0.66 Gy where the average difference is -0.25 Gy. For the percent volume difference at various dose levels for rectum (75 Gy, 70 Gy, 65 Gy, 45 Gy), bladder (80 Gy, 75 Gy, 70 Gy, 65 Gy), and femoral heads (50 Gy), the mean values are within ±0.16%, the standard deviations are less than 0.35%, the minimum values are greater than -0.97%, and the maximum values are less than 0.4%. The difference in mean dose for femoral heads is also negligibly small. All of these numbers indicate that the two dose distributions computed by DL and CS are clinically identical.

Table 1. The Gamma passing rate for DL dose calculation of the eight testing patients.

|  |  | Gamma Index | |
|---|---|---|---|
| Criterion | | 1mm/1% | 2mm/2% |
| Patient | 1 | 99.10% | 100% |
| | 2 | 99.90% | 100% |
| | 3 | 97.60% | 100% |
| | 4 | 99.80% | 100% |
| | 5 | 99.90% | 100% |
| | 6 | 99.90% | 100% |
| | 7 | 95.70% | 99.70% |
| | 8 | 96.30% | 99.80% |
| | Mean | 98.50% | 99.90% |
| | SD | 1.60% | 0.10% |

Table 2. The mean, standard deviation, min, and max of the difference (DL-CS) in IMRT objectives between DL and CS dose calculations for the eight testing patients. The volume differences are in percent volume and the dose differences are in Gy.

|  |  | Mean | SD | Min | Max |
|---|---|---|---|---|---|
| PTV | $D_{95}$ | -0.25 | 0.67 | -1.38 | 0.66 |
| Rectum | $V_{75}$ | -0.16% | 0.32% | -0.78% | 0.27% |
| | $V_{70}$ | -0.16% | 0.35% | -0.97% | 0.10% |
| | $V_{65}$ | -0.09% | 0.20% | -0.46% | 0.14% |
| | $V_{45}$ | -0.13% | 0.13% | -0.33% | 0.08% |
| Bladder | $V_{80}$ | 0.02% | 0.26% | -0.44% | 0.40% |
| | $V_{75}$ | 0.04% | 0.10% | -0.12% | 0.18% |
| | $V_{70}$ | 0.04% | 0.12% | -0.08% | 0.21% |
| | $V_{65}$ | 0.01% | 0.12% | -0.16% | 0.16% |
| Femoral Heads | $D_{mean}$ | -0.02 | 0.13 | -0.24 | 0.12 |
| | $V_{50}$ | 0.00% | 0.00% | 0.00% | 0.00% |

## 4. Conclusions

To our knowledge, this is the first instance of successful dose calculation for radiotherapy using deep learning. It is well known that there is a tradeoff between efficiency and accuracy for all existing dose calculation engines. Our work shows that deep learning-based methods may be able to achieve a high dose calculation accuracy with a high efficiency, which may play an important role for real-time adaptive radiation therapy. Also, since the deep learning-based methods are completely different from the existing measurement-based or model-based methods, they can also be used for secondary dose verification.

5. **Acknowledgements**

We would like to thank the Cancer Prevention and Research Institute of Texas (CPRIT) for providing support through grants IIRACA RP160190 and IIRA RP150485.

## References

1. Ahnesjo A, Aspradakis MM. Dose calculations for external photon beams in radiotherapy. *Phys Med Biol.* 1999;44(11):R99-R155.
2. Luong M-T, Pham H, Manning CD. Effective approaches to attention-based neural machine translation. *arXiv preprint.* 2015;arXiv:1508.04025.
3. Lee J, Cho K, Hofmann T. Fully character-level neural machine translation without explicit segmentation. *arXiv preprint.* 2017;arXiv:1610.03017.
4. Simonyan K, Zisserman A. Very deep convolutional networks for large-scale image recognition. *arXiv preprint.* 2014;arXiv:1409.1556.
5. He K, Zhang X, Ren S, Sun J. Deep residual learning for image recognition. *arXiv preprint.* 2015;arXiv:1512.03385.
6. Ronneberger O, Fischer P, Brox T. U-Net: convolutional networks for biomedical image segmentation. *Medical Image Computing and Computer-Assisted Intervention – MICCAI 2015 Lecture Notes in Computer Science.* 2015;9351:234-241.
7. Nguyen D, Long T, Jia X, et al. Dose prediction with U-net: a feasibility study for predicting dose distributions from contours using deep learning on prostate IMRT patients. *arXiv preprint.* 2017;arXiv:1709.09233. *Scientific Reports.* 2019;9(1):1076.
8. Barragan-Montero AM, Nguyen D, Lu W, et al. Three-dimensional dose prediction for lung IMRT patients with deep neural networks: robust learning from heterogeneous beam configurations. *arXiv preprint.* 2018;arXiv:1812.06934.
9. Nguyen D, Jia X, Sher D, et al. 3D radiotherapy dose prediction on head and neck cancer patients with a hierarchically densely connected U-net deep learning architecture. *arXiv preprint.* 2018; arXiv:1805.10397. *Phys Med Biol.* 2019;64(6).
10. Lu W, Chen M. Fluence-convolution broad-beam (FCBB) dose calculation. *Phys Med Biol.* 2010;55(23):7211-7229.
11. Ahnesjo A. Collapsed cone Convolution of radiant energy for photon dose calculation in heterogeneous media. *Med Phys.* 1989;16(4):577-592.
12. Ulmer W, Kaissl W. The inverse problem of a Gaussian convolution and its application to the finite size of the measurement chambers/detectors in photon and proton dosimetry. *Phys Med Biol.* 2003;48(6):707-727.
13. Ulmer W, Pyyry J, Kaissl W. A 3D photon superposition/convolution algorithm and its foundation on results of Monte Carlo calculations. *Phys Med Biol.* 2005;50(8):1767-1790.
14. Kingma DP, Ba J. Adam: A method for stochastic optimization. *arXiv preprint.* 2014;arXiv:1412.6980.
15. Abadi M, Agarwal A, Barham P, et al. Tensorflow: Large-scale machine learning on heterogeneous distributed systems. *arXiv preprint.* 2016; arXiv:1603.04467.